\documentclass[epjc3]{svjour3}
\smartqed  % flush right qed marks, e.g. at end of proof
\RequirePackage{graphicx}

 \RequirePackage{mathptmx}      % use Times fonts if available on your TeX system
\usepackage{amsmath}
\usepackage{amsfonts}
\usepackage{amssymb}
\usepackage{caption}
\usepackage{subcaption}
\usepackage[font=footnotesize]{subfig}

\RequirePackage{latexsym}
\begin{document}

\title{Dilatonic Scalar Field: A Dynamical System Analysis%\thanksref{t1}
}


\author{Nilanjana Mahata\thanksref{ e1, addr1}
        \and
        Subenoy Chakraborty\thanksref{e2,addr1} %etc.
}
%\email{ nilanjana_mahata@yahoo.com}

%\thankstext{e1}{e-mail: nmahata@math.jdvu.ac.in}
 \thankstext{e1}{e-mail: nmahata@math.jdvu.ac.in}
\thankstext{e2}{e-mail:schakraborty@math.jdvu.ac.in}
\institute{  Department of Mathematics\\
             Jadavpur University\\
             Kolkata- 700032, India\label{addr1}
            % nmahata@math.jdvu.ac.in \\
             %       schakraborty@math.jdvu.ac.in  \label{addr1}
           %\and
        % schakraborty@math.jdvu.ac.in \label{addr2}
}
 %        \and
 %       \emph{Present Address:} if needed\label{addr3}
\maketitle

\begin{abstract}
The work deals with homogeneous and isotropic, flat FRW model of the universe which is filled up with non-interacting dark matter and dark energy to compatible with recent observational evidences. By choosing the dark energy in the form of a dilatonic scalar field, the evolution equations  are reduced to an autonomous system. A phase space analysis is done around the critical points and stability criteria is examined. Finally, cosmological implications of the nature of the critical points are discussed.

\end{abstract}

\section{Introduction}
%\label{intro}
Recent cosmological observations   [1-3] strongly indicate an accelerated expanding universe with flat spatial geometry. The challenging issue of incorporating this accelerating phase in the framework of standard cosmology is taken care of by introducing an unknown energy component , dubbed as dark energy having marginally dominant negative pressure . Usually, the vacuum energy  is the natural choice of dark energy. Cosmological constant is avoided due to at least a couple of embarrassing issues namely i) the 'fine-tuning' problem (why the vacuum energy is so small in particle physics units) and ii) the 'coincidence problem'(though generically 'small', the cosmological constant turns out to be exactly of the value required to become dominant at present). To overcome these puzzles, it is normally assumed that  vacuum energy is balanced by some unknown cancellation mechanism and in turn there is a dark energy component having variable equation of state. There are several candidates in the literature  for dynamical dark energy namely quintessence  [4], K-essence  [5], tachyons  [6],  braneworld  [7], chaplygin gas  [8], dilaton  [9] etc. From the motivation of the supersymmetric field theories and string/M theory, the scalar field may be of dilatonic type. Also dilatonic scalar field eliminates some quantum instabilities with respect to the phantom field models of dark energy [10]. Further it has been shown  [10] that the coupling of  the dilatonic field  to other dark sector(dark matter) may lead to a final cosmological attractor with both an accelerated expansion and a constant ratio between dark matter and dilatonic energy densities. This scaling solutions resolves the 'coincidence problem' partially and is consistent with recent Type Ia supernova data  [11].

 In the present work, we analyze the possible cosmological behaviour of the dilatonic scalar field in FRW space time. By defining auxiliary variables, the field equations are reduced to an autonomous system. Then by performing a phase- space analysis and stability criteria , we study the possible late time solutions and relevant cosmological parameters. The plan of the paper is as follows: in the following section we present the basic equations for dilaton scalar field related to FRW cosmology. Section 3 deals with dynamical system and phase space analysis, stability criteria is discussed in section 4. Finally, cosmological implications and conclusions are presented in section 5. \\
% {Section title}
%\label{sec:1}
%Text with citations \cite{RefB} and \cite{RefJ}.
\section { Basic Equations for Dilatonic Scalar Field: FRW cosmology}

The Lagrangian density of the dilatonic dark energy is chosen as the pressure density of the scalar field as [9]
\begin{equation}
{\L}_{\phi} = p_{\phi}= -X + \alpha e^{\lambda\phi}X^{2}
\end{equation}

where $ \alpha$ and $\lambda$ are positive constants and $  X = \dot{\phi}^{2}/2 $. As a consequence, the energy density has the expression

%\label{sec:2}
%as required. Don't forget to give each section
%and subsection a unique label (see Sect.~\ref{sec:1}).
%\paragraph{Paragraph headings} Use paragraph headings as needed.
\begin{equation}
\rho_{\phi} = - X + 3\alpha e^{\lambda\phi}X^{2}
\end{equation}
 So the equation of state parameter of the dilaton DE has the form\\
 \begin{equation}
   \omega_{d}= \frac{p_{\phi}}{\rho_{\phi}} = \frac{- 1 + \alpha e^{\lambda\phi}X}{- 1 + 3\alpha e^{\lambda\phi}X}
\end{equation}
Also, the density parameter for the dilatonic scalar field (representing the dark energy) and the total effective equation of state parameter $(w_{t})$ are respectively given by\\
\begin{equation}
\Omega_{\phi} = \frac{\rho_{\phi}}{3H^{2}} = \frac{X}{3H^{2}}(3\alpha e^{\lambda\phi}X - 1)
\end{equation}
and
\begin{equation}
 w_{t} = \frac{p_{\phi}}{\rho_{\phi} + {\rho_{m}}}= w_{d}\Omega_{\phi} = \frac{X}{3H^{2}}(\alpha e^{\lambda\phi}X - 1)
\end{equation}
The effective sound speed is given by\\
\begin{equation}
   Cs^{2}= \frac{\partial p_{\phi}/\partial X}{\partial\rho_{\phi}/\partial X} = \frac{- 1 + 2\alpha e^{\lambda\phi}X}{- 1 + 6\alpha e^{\lambda\phi}X}
\end{equation}
which can be expressed in terms of the equation of state parameter $w_{d}$ as\\
\begin{equation}
   Cs^{2} = \frac{1 + w_{d}}{5-3w_{d}}
\end{equation}

It is to be noted that the above definition of the sound speed is due to the evolution of linear adiabatic perturbations in a scalar field dominated universe. For homogeneous and isotropic flat FRW model of the universe, the equation of motion of the scalar field has the form\\
\begin{equation}
   \frac{d}{dt}[\frac{\partial {\L}_{\phi}}{\partial X}\dot{\phi}] + 3H \frac{\partial \L_{\phi}}{\partial X}\dot{\phi} + \frac{\partial {\L}_{\phi}}{\partial \phi} = 0
\end{equation}
or explicitly,
%\begin{equation} % to write in two lines
%\begin{split}
 %  (6\alpha e^{\lambda\phi}X - 1 )+ 2\lambda\alpha e^{\lambda\phi}X \dot{\phi}^{2} + 3H\dot{\phi}(2\alpha %e^{\lambda\phi}X - 1 )\\
 %+ \alpha \lambda e^{\lambda\phi}X^{2} & = 0
%\end{split}
\begin{equation}
%\begin{split}
   (6\alpha e^{\lambda\phi}X - 1 )+ 2\lambda\alpha e^{\lambda\phi}X \dot{\phi}^{2} + 3H\dot{\phi}(2\alpha e^{\lambda\phi}X - 1 )\\
 + \alpha \lambda e^{\lambda\phi}X^{2}  = 0
%\end{split}
\end{equation}
with H as the usual Hubble parameter. Now for flat FRW model, the Friedman equations are (choosing $ 8\pi G = c = 1 $)\\
\begin{equation}
 H^{2} = { \frac{1}{3}}(\rho_{m} + \rho_{\phi}),
\end{equation}
\begin{equation}
   \dot{H} = - \frac{1}{2}(\rho_{m} + \rho_{\phi} + p_{\phi}),
\end{equation}
where dark matter (DM) is in the form of dust of energy density $\rho_{m}$ and $(\rho_{\phi}, p_{\phi})$ are the energy density and thermodynamic pressure  for the dark energy (DE). As at present the universe is dominated by DM and DE so we have not taken into consideration of the baryonic matter and radiation for simplicity.
\section{Dynamical System and Critical Points: Phase Space Analysis}

As the evolution equations are complicated so to make a qualitative analysis we transform the cosmological evolution equations into an autonomous dynamical system by introducing auxiliary variables x and y and we have the self autonomous system:\\
\begin{equation}
   \overrightarrow{X}' = \overrightarrow{f}(\overrightarrow{X})
\end{equation}
Here the column vector $ \overrightarrow{X}$ is constituted by the auxiliary variables i.e $ \overrightarrow{X} = \left (\begin{array}{c}
x\\
y \end{array}\right)$ and $ \overrightarrow{f}(\overrightarrow{X})$ is the column vector with the r.h.s of the autonomous system and prime denotes differentiation with respect to $\tau = \ln a $. The critical points $ \overrightarrow{X_{c}}$ are obtained from $ \overrightarrow{X}' = 0$ i.e $\overrightarrow{f}(\overrightarrow{X_{c}}) = 0 .$
The stability criteria of a critical point is obtained by the perturbative expansion about the critical point i.e by setting $ \overrightarrow{X} = \overrightarrow{X_{c}} + \overrightarrow{U}$. Here the perturbation $\overrightarrow{U}$ can be obtained from the relation\\
\begin{equation}
 \overrightarrow{U}' = M\overrightarrow{U}
\end{equation}
where M is a $2\times 2 $ matrix, containing the perturbation coefficients. Then the eigen values of M at the critical point will characterize the type of critical point as well as the nature of it. In fact, if $ Tr M < 0 $ and $ det M > 0 $ , the critical point is said to be a stable point.\\
In the present problem the auxiliary variables are chosen as
\begin{equation}
 x = \phi'     ,   y = \alpha e ^{\lambda\phi}X
\end{equation}
Then various cosmological parameters and the sound speed can be expressed in terms of these auxiliary variables as\\
 \begin{equation}
%\begin{split}
  w_{d}  = \frac{y - 1}{3y - 1}  ,  \Omega_{\phi}= \frac{x^{2}}{6}(3y - 1) ,\\
  w_{t}  = \frac{x^{2}}{6}(y - 1) ,   Cs^{2}  = \frac{2y - 1}{6y - 1}
%\end{split}
 \end{equation}
As $  \Omega_{\phi}\in [ 0,1] $ so the auxiliary variables are restricted by the relation
\begin{equation}
  0  \leq  \frac{x^{2}(3y - 1)}{6}  \leq  1
\end{equation}
The region given by inequality (16) has two disjoint open infinite region described by the two branches of the curve $ {x^{2}(3y - 1)} = 6 $ and the asymptotes $ x = 0, y = \frac{1}{3}$ as shown in figure 1.\\

%\begin{figure}
 % \includegraphics{dila1.eps}
%\caption{ Plot  of the curve $ {x^{2}(3y - 1)} = 6 $}
 %\end{figure}
\begin{figure}
\centering
\begin{minipage}{.5\textwidth}
  %\centering
  \includegraphics[width=1.0\linewidth]{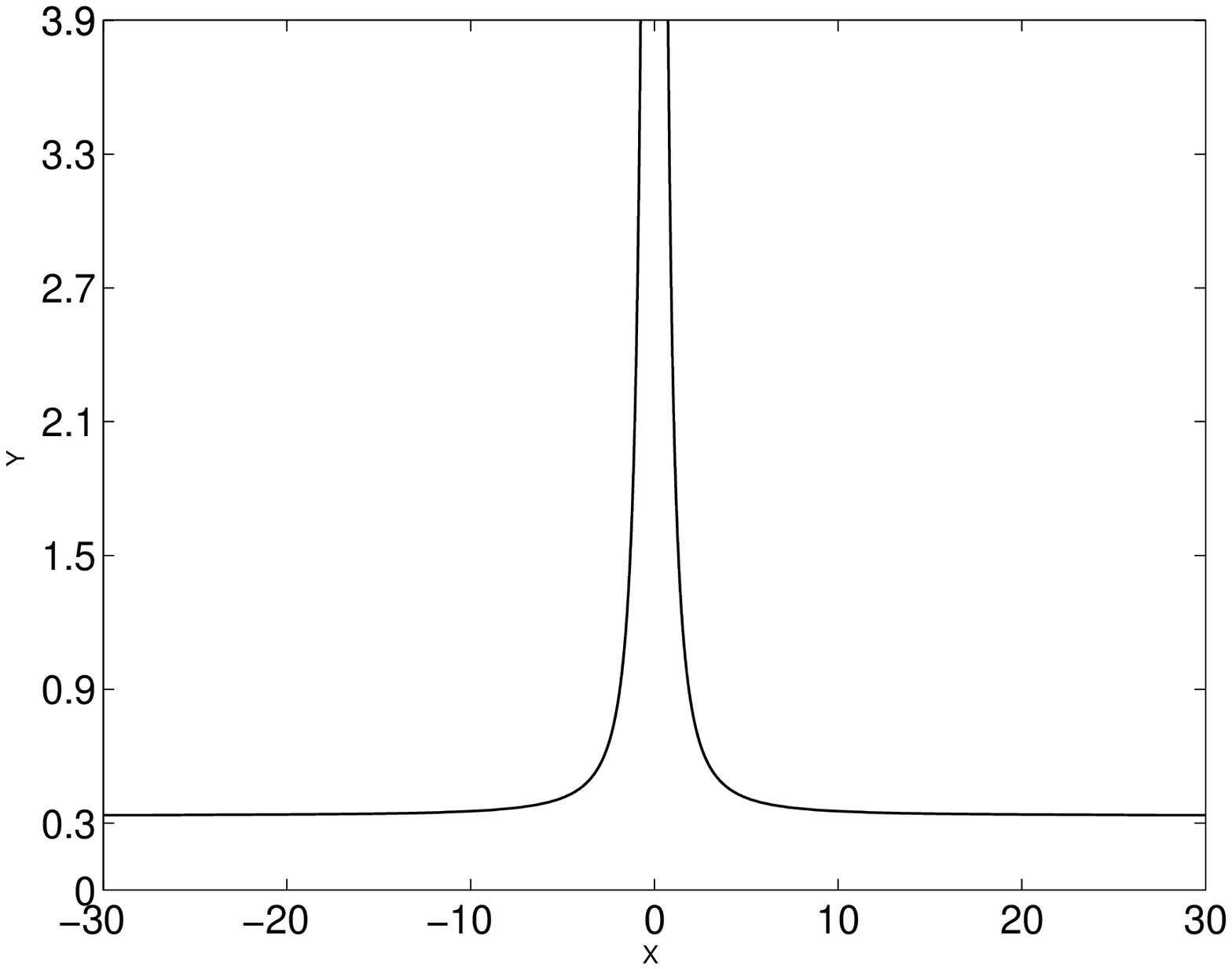}
  \caption{Plot  of the curve $ {x^{2}(3y - 1)} = 6 $}
 % \label{fig:test1}
\end{minipage}%
%\hspace{0.5cm}
\begin{minipage}{.5\textwidth}
 %\centering 
  \includegraphics[width= 1.0 \linewidth]{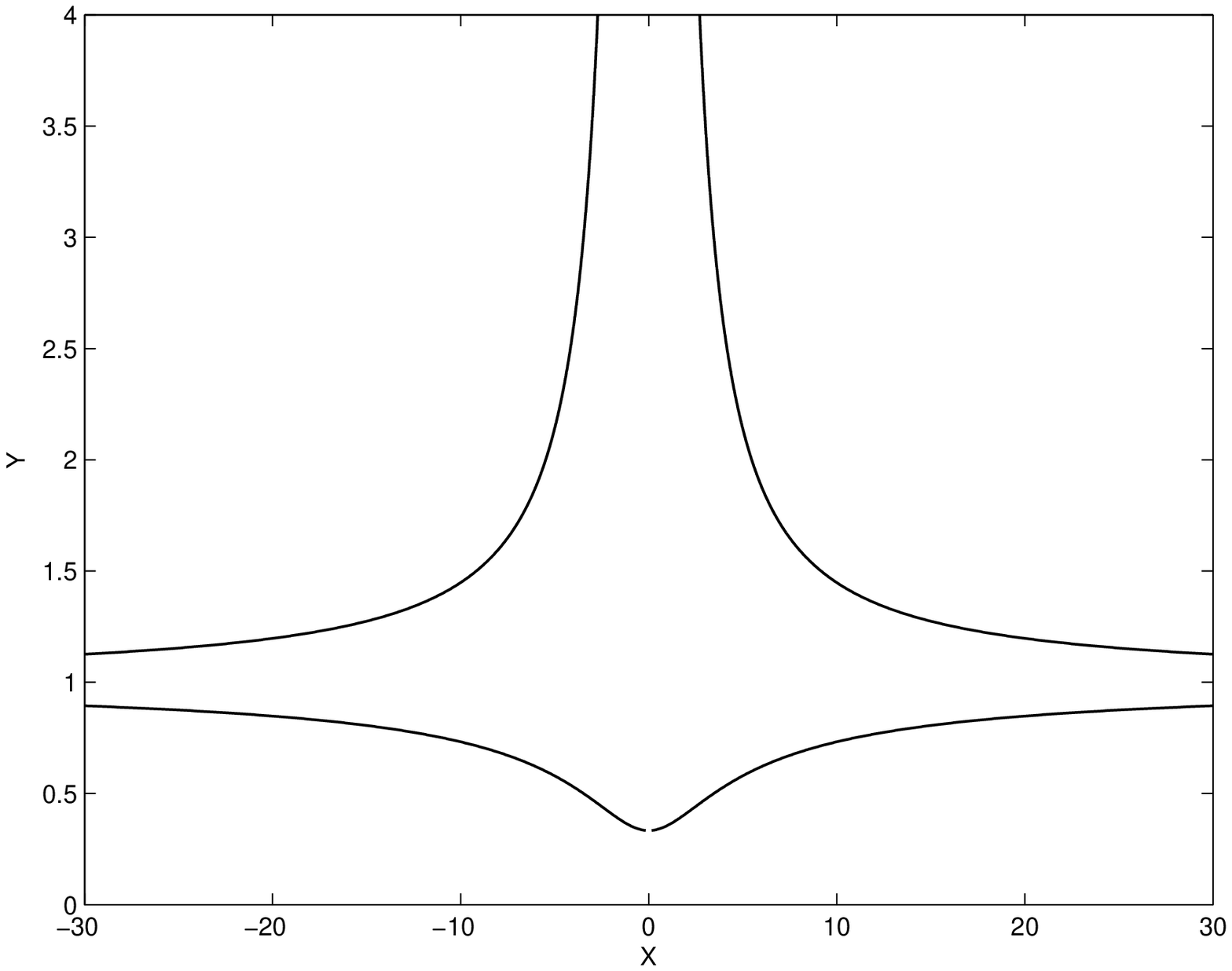}
\centering
  \caption{The curve $ x^{2} = 6(3y - 1 )/(y - 1 )^{2}$}
 % \label{fig:test2}
\end{minipage}
\end{figure}

Normally the dynamical system is analyzed not only in the finite part of the phase space but also at the points at infinity considering Poincare projection method  [12]. Now for the above choice of the auxiliary variables the field equations (9) -(11) reduce to an autonomous system as\\
%\begin{eqnarray}
%\begin{equation}
%\begin{split}
 % x'= \frac{x}{2(6y - 1)} [3(6y - 1)\{1 + \frac{x^{2}(y - 1)}{6}\}\\
 %-5\lambda xy - 6(2y - 1)]
%\end{split}
%\end{equation}
%\begin{equation}
\begin{eqnarray}
  x'= \frac{x}{2(6y - 1)} [3(6y - 1)\{1 + \frac{x^{2}(y - 1)}{6}\} -5\lambda xy - 6(2y - 1)]\\
%\end{equation}
%\begin{equation}
  y' = \frac{y}{6y - 1}[\lambda x(y - 1) - 6(2y - 1 )]
%\end{equation}
\end{eqnarray}
Note that this self autonomous system is valid in the whole phase plane ( not only at critical points) except at $ y \neq \frac{1}{6}$ . Now the critical points are obtained by solving the algebraic equations which are formed by equating to zero the r.h.s of equation (17)- (18) and we have immediately seen that $(0,0)$ and $(0,\frac{1}{2})$ are the critical points of which $(0,\frac{1}{2})$  is in the admissible region (fig. 1).The other critical points are obtained by solving the following two algebraic equations\\
\begin{eqnarray}
  x^{2}(6y - 1)(y - 1)- 10\lambda xy + 6(2y + 1) = 0\\
  \lambda x(y - 1) - 6(2y - 1 ) = 0
\end{eqnarray}

Thus for different values of $ \lambda $ we have different critical points which may or may not be in the physically admissible region. As $ \lambda $ becomes vanishingly small we have two critical points at $( \pm 2\sqrt{3}, \frac{1}{2})$ . From the inequation (16), we see that the physically allowed region in the phase plane lies in the half plane $  y  \geq  \frac{1}{3}  $  [see fig 1 also]. Further the critical point will have negative  x co-ordinate if it lies in the horizontal strip  $ \frac{1}{2}  <  y  <  1 $ , while the critical points will have positive  x co-ordinate in the horizontal strip  $  \frac{1}{3} <  y  <  \frac{1}{2}$ and in the half plane $ y > 1 . $ Now eliminating $ \lambda $  between the two equations (19) and (20),  we find that the critical points will lie on the curve $ x^{2} = 6(3y - 1 )/(y - 1 )^{2}$ which has asymptotes along $ x = 0 $ and $ y = 1 $   [see fig 2].  The curve of critical points $ x^{2} = 6(3y - 1 )/(y - 1 )^{2}$  and the curve $ {x^{2}(3y - 1)} = 6 $  intersect at $( \pm 2\sqrt{3}, \frac{1}{2}) $ [ see fig 3].

\begin{figure}
\centering
  \includegraphics[width=0.50\textwidth]{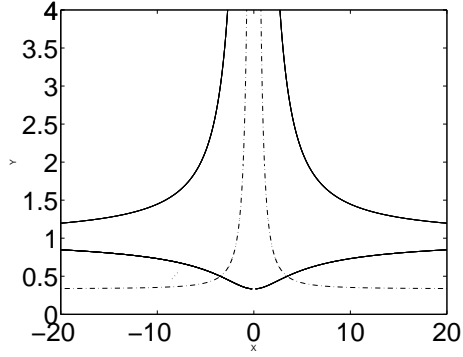}
\caption{The curve of critical points $ x^{2} = 6(3y - 1 )/(y - 1 )^{2} $ and the admissible region given by the curve $ {x^{2}(3y - 1)} = 6 $ are shown here. The curves intersect at the critical point
 $ (\pm2\sqrt{3},\frac{1}{2})$.  }
%label{fig:1}       % Give a unique label
\end{figure}
\begin{figure}
\centering
  \includegraphics[width=1.0\textwidth]{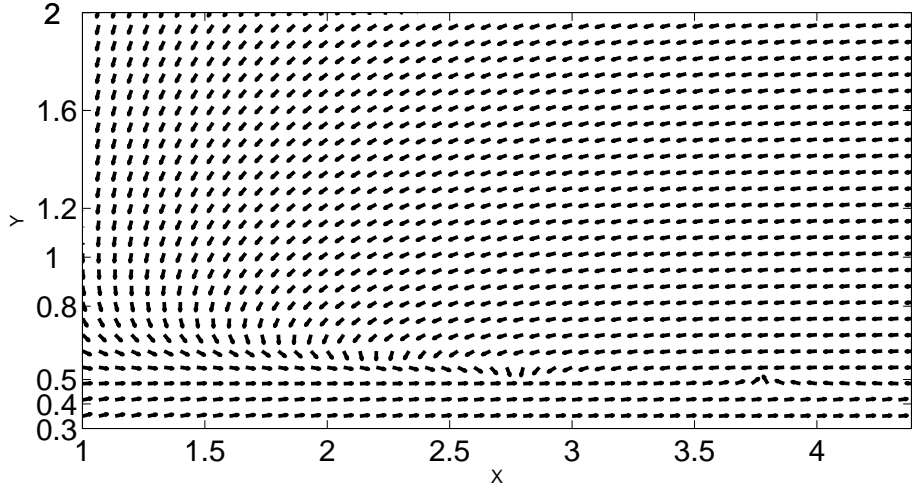}
\caption{Direction field  for the system given by (17)-(18) near the critical point
 $ (2\sqrt{3},\frac{1}{2})$  }
%label{fig:1}       % Give a unique label
\end{figure}

 However, for $ x_{c}  \neq  0 $, and $ y_{c}  \neq  0 $, the relation between the coupling parameter $ \lambda$ and $ y_{c}$  is  $  \lambda = \frac{\sqrt{6}\mid 2y_{c} - 1 \mid}{\sqrt{3y_{c} - 1}}$ .
Further for cosmic acceleration, we have  $ -1  \leq  w_{d} \leq  - \frac{1}{3}  $   i.e    $\frac{1}{2}  \leq  y  \leq  \frac{2}{3} $ and as a result from equation (15),  $C_{s}$ is restricted to the range $ 0  \leq  C_{s} \leq  \frac{1}{3} $ . Thus the present model does not violate causality.  However, for phantom era ( i.e $ w_{d} < -1 $),  y is restricted to the strip $ \frac{1}{3} < y  < \frac{1}{2} $ but in this case sound speed become imaginary.\\

Table I shows the possible critical points for the autonomous dynamical system (17)-(18) and the values of the different  cosmological parameters (relevant to the present context) at the critical points. There are numerous critical points on $ x^{2} = 6(3y - 1 )/(y - 1 )^{2} $, but remembering \emph{} the admissible region [given in fig 1] and validity constraints on $ \lambda $,  the critical points  having  positive x co-ordinates and $ \frac{1}{3} < y  < \frac{1}{2} $ are only considered [ see fig 3]. Critical points $  C_{4} $  and  $ C_{5} $ [see Table 1] are two of them.  In the following we shall analyze the above critical points in details.\\

 Critical point $C_{1}$ : $ x_{c_{1}} = 0 , y_{c_{1}}= \frac{1}{2} $ \\

At this critical point $ \Omega_{\phi} = 0 $  so  the universe is completely dominated by dark matter . The dark energy in the form of the dilatonic scalar field behaves as a cosmological constant $ (w_{d} = -1 )$\\

 Critical point $C_{2}$ : $ x_{c_{2}} = 2\sqrt{3} , y_{c_{2}}= \frac{1}{2} $\\

This critical point corresponds to the limiting situation $\lambda \rightarrow 0 $ . It is completely dominated by dilatonic scalar field.The scalar field and the resulting fluid are both in the phantom crossing. \\
%\begin{table*}
\begin{table}
\caption{ Values of different cosmological parameters at critical points}
%\label{tab:1}       % Give
\begin{tabular}{|c|c|c|c|c|c|c|c|}
 \hline
  Critical point & Nature & Eigen values & $C_{s}^{2}$ & $w_{d}$ & $ \Omega_{\phi}$ & $ w_{t} $ & $ \lambda $ \\\hline
 $ C_{1}(0, \frac{1}{2})$  & Saddle point & $  \frac{3}{2},-3 $& 0 & -1 & 0 & 0 & 0 \\\hline
 $ C_{2}(2\sqrt{3}, \frac{1}{2})$ & stable node & $ -3, -3 $ & 0 & -1 & 1 & -1 & 0 \\\hline
 $  C_{3}(-2\sqrt{3}, \frac{1}{2})$  & stable node & $ -3, -3 $ & 0 & -1 & 1 & -1 & 0\\\hline
 $ C_{4}(\sqrt{\frac{10}{3}}, \frac{2}{5})$ & Stable node & $ \frac{-\frac{37}{7} \pm \frac{\sqrt{185}}{7}}{2}$ & $-\frac{1}{7}$ & -3 & $\frac{1}{9}$ & $-\frac{1}{3}$ & $\sqrt{\frac{6}{5}} $ \\\hline
 $ C_{5}(\frac{6\sqrt{6}}{7}, \frac{5}{12})$  & stable node & $ \frac{-\frac{112}{21} \pm \frac{\sqrt{172}}{63}}{2}$ & $-\frac{1}{9}$ & $-\frac{7}{3}$ & $\frac{9}{49}$ & $-\frac{3}{7}$ & $ \sqrt{\frac{2}{3}}$ \\\hline

\end{tabular}
\end{table}
 %\end{table*}

 Critical point $C_{3}$ : $ x_{c_{3}} = - 2\sqrt{3}, y_{c_{3}}= \frac{1}{2} $ \\

This critical point  also corresponds to the limiting situation $\lambda \rightarrow 0 $ and has identical behaviour as $C_{2}$ . The values of the cosmological parameters are presented in table I.

 Critical point $C_{4}$ : $ x_{c_{4}} = \sqrt{\frac{10}{3}} , y_{c_{4}}= \frac{2}{5} $ \\

This point lies on the line of critical points shown in figure 2. For this stable node the coupling parameter $\lambda$ is non-zero but sound speed becomes imaginary. Here the combined 2-fluid system is in the quintessence barrier and is dominated by the dark matter.\\

 Critical point $C_{5}$ : $ x_{c_{5}} = \frac{6\sqrt{6}}{7} , y_{c_{5}}= \frac{5}{12} $ \\

Similar to the previous one this critical point also lies on the line of critical points. Here the resulting fluid is well within the quintessence era and is also dominated by the dark matter component.

\section{ Stability Analysis}

 To study the stability of the present model , we first note that stability of  the critical points do not imply the model to be stable . In this section , we examine the stability of the model both classically and quantum mechanically . In general, to address the quantum stability of a scalar field $ \phi $ , one has to consider the dynamics of the small fluctuations $ \delta\phi $ around a background value $ \phi_{0}$ which is a solution of classical equations such that $ \phi_{0} \neq 0 $ and $ \dot{\phi_{0}} \neq 0 $. Thus one may write
                  $ \phi (t,\overrightarrow{r}) = \phi_{0}(t) + \delta \phi(t,\overrightarrow{r}) $.\\
i.e the scalar field is split up into a homogeneous part and a fluctuation $ \delta \phi $ . As mostly we deal with UV instabilities so it is not essential to choose Minkowaski background metric rather the choice to be such that, at least locally , there is a time direction to have the above decomposition of $ \phi $ . Then the Hamiltonian  [9,13] for the fluctuations (upto second order) can be written as\\

$ H = (P_{X} + 2XP_{XX})\frac{(\delta\dot{\phi})^{2}}{2} + P_{X}\frac{(\nabla \delta\phi)^{2}}{2} - P_{\phi\phi}\frac{(\delta\phi)^{2}}{2}$,\\

where P is the Lagrangian density for the scalar field $ \phi $. Now for positive definiteness of H, we have the following restrictions:\\

$ a \equiv  P_{X} + 2XP_{XX} \geq 0 $ , $ b  \equiv P _{X} \geq 0 , $ $  c = - P_{\phi\phi} \geq 0 $ \\

Here a and b are related to classical stability through the speed of sound as \\

  $          C_{s}^{2} \equiv  \frac{P_{X}}{\rho_{X}} = \frac{b}{a} $\\

and it appears in cosmological perturbations as a coefficient of the term $(momentum/scale factor)^2 $ . Thus for the stability of the model, we have the following criteria :\\

a) $ C_{s}^{2}\geq 0 $ (classical fluctuation),  
b) $ a > 0 , b  \geq 0 $ (quantum phenomena)\\

In the present problem we have $ p = - X + \alpha e^{\lambda\phi}X^{2}$, so $  a = 6y - 1$  ,  $ b = 2y - 1 $ and hence $ C_{s}^{2} = \frac{6y - 1}{2y - 1} $.\\

So for classical stability we have $ y \geq \frac{1}{2} $ or $ y < \frac{1}{6} $ while for quantum stability we have $ y \geq \frac{1}{2} $  . Thus all stable critical points lie in the half plane $ y \geq \frac{1}{2} $. As $ y < \frac{1}{2 } $ corresponds to phantom domain while the horizontal strip $ \frac{1}{2} < y  < \frac{2}{3} $ represents the quintessence era,  so one may conclude that the present DE model is stable both classically and quantum mechanically in the quintessence domain while the phantom region is unstable in all respect.
\section{Cosmological Implications and Conclusion}

We shall now analyze the critical points from the perspective of cosmology based on the above phase space analysis and stability criteria.\\

From table I we see that for the first three critical points the model is both classically and quantum mechanically stable while the model is unstable for the last two critical points. However , except the first one, all the critical points are stable one. So the critical point $C_{1}$ is not relevant from cosmological view point. The first three critical points correspond to phantom  barrier of the DE fluid while last two are in the phantom region corresponding to DE fluid. Thus all the critical points represent late time accelerated expansion at present epoch.But from the view point of stability of the system only critical points $ C_{2}$ and $ C_{3}$ are of interest in the perspective of cosmological scenario. It should be noted that critical points $ C_{2}$ and $ C_{3}$ are the limiting cases corresponding to $ \lambda \rightarrow 0 $. Therefore from the phase space analysis one may conclude that dilatonic scalar field can be considered as a possible candidate for DE to exhibit present accelerating phase of the universe.\\\\\\

\begin{acknowledgements}

One  of  the  author ( Subenoy Chakraborty ) is  thankful  to  UGC-DRS  Programme,  Dept. of Mathematics, Jadavpur University.

\end{acknowledgements}

\end{document}